\documentclass{aa}
\usepackage{amsmath}
\usepackage[shortlabels]{enumitem}
\usepackage{amsmath,amssymb}
\usepackage{xcolor}
\usepackage{graphicx}
\usepackage[export]{adjustbox}
\usepackage{mathtools}
\usepackage{amsmath}
\usepackage{booktabs}
\usepackage{bm}

\usepackage{graphicx}
\usepackage{tabularx}
\usepackage{txfonts}

\begin{document} 

   \title{Impact of the measured parameters of exoplanets on the inferred internal structure}

   \author{J.F. Otegi\inst{1,2} 
          \and
          C. Dorn\inst{1}
          \and
          R. Helled\inst{1}
          \and
          F. Bouchy\inst{2}
          \and
          J. Haldemann\inst{3}
          \and
          Y. Alibert\inst{3}
          }

   \institute{Institute for Computational Science, University of Zurich,
              Winterthurerstr. 190, CH-8057 Zurich, Switzerland\\
              \email{jon.fernandezotegi@unige.ch}
         \and
              Observatoire Astronomique de l’Universit\'e de Gen\`eve, 51 Ch. des Maillettes, – Sauverny – 1290 Versoix, Switzerland\\
         \and
              Departement of Space Research \& Planetary Sciences, University of Bern, Gesellschaftsstrasse 6, CH-3012 Bern, Switzerland
             }


  \abstract
   {Exoplanet characterization is one of the main foci of current exoplanetary science. For super-Earths and sub-Neptunes, we mostly rely on mass and radius measurements, which allow us to derive the mean density of the body and give a rough estimate of the bulk composition of the planet. However, the determination of planetary interiors is a very challenging task. In addition to the uncertainty in the observed fundamental parameters, theoretical models are limited owing to the degeneracy in determining the planetary composition.}
   {We aim to study several aspects that affect the internal characterization of super-Earths and sub-Neptunes: observational uncertainties, location on the M-R diagram, impact of additional constraints such as bulk abundances or irradiation, and model assumptions.}
   {We used a full probabilistic Bayesian inference analysis that accounts for observational and model uncertainties. We employed a nested sampling scheme to efficiently produce the posterior probability distributions for all the planetary structural parameter of interest. We included a structural model based on self-consistent thermodynamics of core, mantle, high-pressure ice, liquid water, and H-He envelope.  }
   { Regarding the effect of mass and radius uncertainties on the determination of the internal structure, we find three different regimes: below the Earth-like composition line and above the pure-water composition line smaller observational uncertainties lead to better determination of the core and atmosphere mass, respectively; and between these regimes internal structure characterization only weakly depends on the observational uncertainties. We also find that using the stellar Fe/Si and Mg/Si abundances as a proxy for the bulk planetary abundances does not always provide additional constraints on the internal structure. Finally we show that small variations in the temperature or entropy profiles lead to radius variations that are comparable to the observational uncertainty. This suggests that uncertainties linked to model assumptions can eventually become more relevant to determine the internal structure than observational uncertainties. }
   
   \vspace{7mm}
   {}

   \maketitle
%

\section{Introduction}

Over the past few years, the characterization of planet interiors has been the subject of extensive research. The large number and diversity of discovered exoplanets has allowed us to identify multiple planet populations. Among these, there is an increasing interest in super-Earths and sub-Neptunes, which cover the transition from terrestrial planets to gas giants and have no analog in our solar system. Major improvements in observational techniques allow for relatively precise measurements of mass and radius. The precision of the planetary radius is limited by the uncertainty of the stellar size, since the transit depth scales as $R_P^2/R_*^2$. Recently \cite{Berger18} presented revised radii of more than 180 000  Kepler stars, leading to a remarkable improvement of the median radius precision.  In some cases, space missions can perform high precision photometry and asteroseismology and can reach relative radius uncertainties of about 3\% \citep[e.g.,][]{Hatzes16}. In addition, the current most advanced spectrographs have radial velocity precision of ~1m/s. This precision was recently improved with instruments such as ESPRESSO (Echelle Spectrograph for Rocky Exoplanets and Stable Spectroscopic Observations; e.g., \cite{Pepe2018} and references therein), which is expected to have an accuracy close to 10 cm/s. Therefore, a significant improvement in the mass determination is also expected, allowing us to reach a relative uncertainty better than 10\%.

The masses and radii can be used to estimate the interior structure and composition of a planet. However, determining the internal structure is extremely challenging as a consequence of intrinsic degeneracy because several compositions can lead to identical mass and radius \cite[e.g.,][]{rogers10,Lopez14,Dorn15,Dorn17, Lozovsky18}. Furthermore, for a planet of given mass and composition, the radius depends on several aspects such as the choice of equation of state (EOS), the envelope structure (differentiated, fully mixed, or with a compositional gradient), or the temperature.  This degeneracy is critical owing to the large number of free parameters needed to model the interior of an exoplanet and the few observational constraints. In order to determine how well one interior model compares with the other possible models that also fit the data and which structural parameters can be constrained, \cite{Dorn17} presented a generalized Bayesian inference method to quantify the degeneracy and correlation of the planetary structural parameters. \\

In this work we explore the limitations of constraining the internal structure of super-Earths and sub-Neptunes (focusing on planets with masses up to 25M$_{\oplus}$ and radii up to 3.5M$_{\oplus}$). We use a Bayesian inference analysis together with a nested dampling technique \cite[e.g.,][]{Skilling04} to discuss several aspects that affect interior characterization: observational uncertainties, location in the mass-radius (M-R) diagram, additional observational constraints as bulk abundances, masses derived from published M-R relationships for fixed radii, and the uncertainty related to model assumptions. First, we study the influence of the data uncertainty on the determination of the internal structure. In \cite{Dorn15}, they quantified the information gained by higher data precision. This study aims to show systematically the effect of data uncertainty for a wide range of masses and bulk densities ($0.5-14.7 \ g.cm^{-3} $).  Additional constraints are crucial to reduce the degeneracy, and the assumption that relative abundances of refractory elements (e.g., Fe/Si, Mg/Si) of a planet are similar to that of its host star has been proposed to reduce the existing degeneracy \cite[e.g.,][]{Grasset09,Dorn15}. Several solar system and planet formation studies have stated that there is a direct correlation between stellar and planetary relative bulk abundances \cite[e.g.,][]{Carter12,Lodders13,Righter02,McDonough95,Bond10,Elser12,Johnson12,Thiabaud15,Wang18}. We explore under what conditions interior estimates can be improved by constraints on planetary bulk abundances taken from stellar proxies. \\

Most of the discovered exoplanets do not have measured masses and radii, and published M-R relationships allow us to estimate the mass for a given radius and vice versa. In addition, M-R relationships describe the main properties of various classes of exoplanets. Several studies have been dedicated to the investigation of the M-R relationship of observed exoplanets. The M-R relationships are power laws of the type $M=A R^B$ and are based on exoplanet data \citep[e.g.,][]{Weiss13,Weiss14,Wolfgang15, Bashi17}. Recently, \cite{Zeng16} inferred a semiempirical M-R relationship depending on the core mass fraction, followed by a detailed forecasting model using a probabilistic M-R relationship via Markov chain Monte Carlo (MCMC) \citep[e.g.,][]{Chen17}. In addition, in \cite{Otegi2019} we presented an updated exoplanet catalog based on reliable, robust, and as much as possible accurate mass and radius measurements of transiting planets up to 120 $M_{\oplus}$, and we inferred two new empirical M-R relationships corresponding to rocky and volatile-rich populations.  We study inferred interior parameters using the mass calculated from above-mentioned M-R relationships. 

Finally, we study how variations in the temperature profiles lead to radius uncertainties comparable to observational uncertainties. In this work, we assess the importance of the various sources of uncertainty related to model assumptions and study the effect of variations of the temperature profile on the planetary radius, which has not been explored yet. 

\vspace{10mm}

\section{Method}

\subsection{Synthetic sample}

We studied the internal structure of a sample of 20 synthetic planets with different masses and radii up to $25 M_{\oplus}$ and $3.5 R_{\oplus}$ (listed in Table 1). We aim to better understand the transition between rocky and volatile-rich exoplanets. \cite{Fulton2017} found that there is a lack of planets with radii between $1.5 R_{\oplus}$ and $2 R_{\oplus}$, known as "evaporation valley" \cite[e.g.,][]{Owen2013,Jin14,Lopez14}, suggesting a transition between the super-Earth and sub-Neptune populations. In addition, in \cite{Otegi2019} we did a careful analysis to build an exoplanet catalog as reliable as possible and we found a transition region from rocky to volatile-rich exoplanets, which corresponds to a mass in the range 5-25 $M_{\oplus}$, and a radius in the range 2-3 $R_{\oplus}$. The M-R range covered by the synthetic planets includes the transition between these two populations. Furthermore, more massive planets may have massive atmospheres in which the effects of electron degeneracy pressure are significant, and these are not accounted for in our atmospheric model.   \\ 

 \begin{figure}[h]
  \begin{center}
    \includegraphics[scale=0.4]{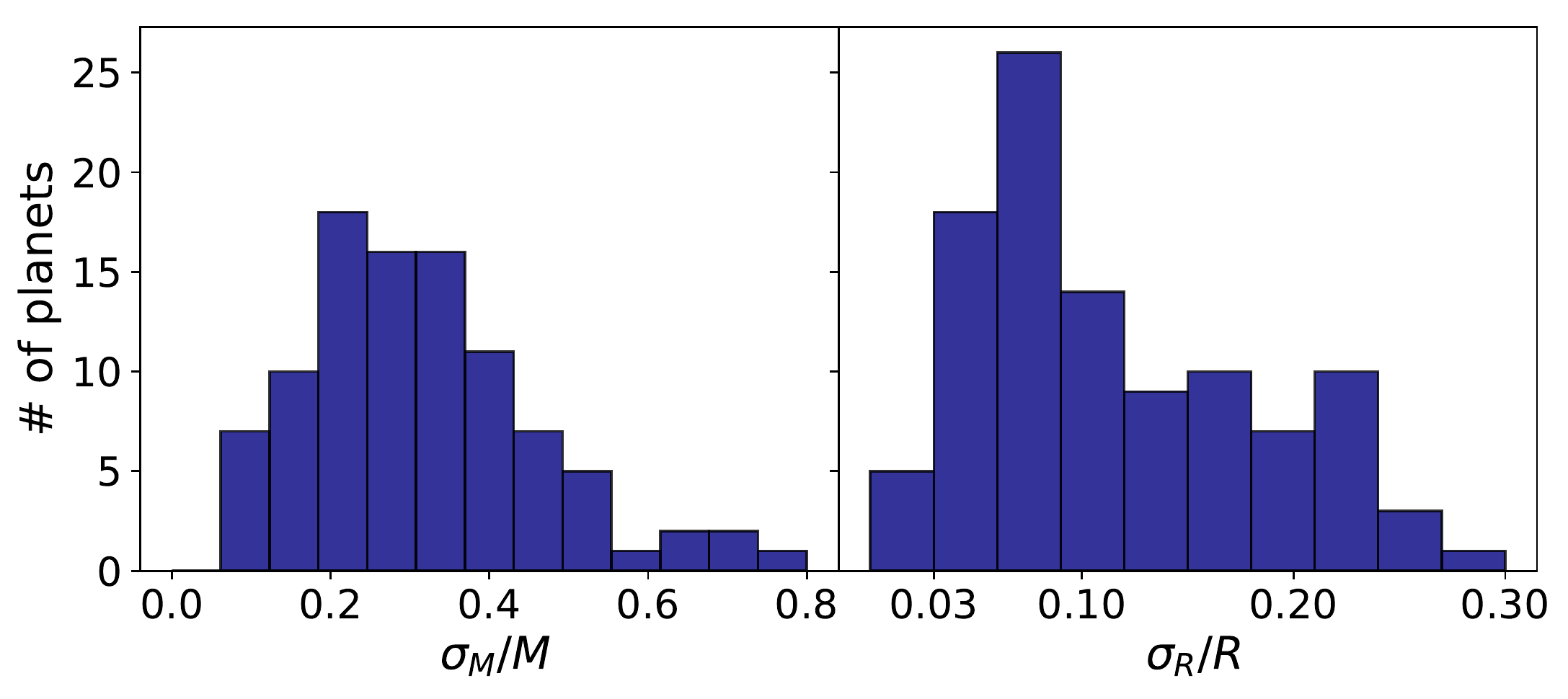}
    \caption{Distributions of the relative mass (left) and radius (right) uncertainties for the 115 observed exoplanets less massive than $25 M_{\oplus}$ from the NASA Exoplanet Archive. }
 
  \end{center}
\end{figure}

The 20 synthetic planets for our sample were chosen to represent the properties of the exoplanet catalog presented in \cite{Otegi2019}, in which we presented a revisited exoplanet catalog based on robust, reliable, and precise mass and radius measurements for transiting exoplanets (with $\sigma_M / M = 25 \%$ and $\sigma_R / R=8 \%$). This exoplanet catalog is dominated by exoplanets for which the masses have been measured through radial velocity, so it is dominated by relatively short-period exoplanets. Exoplanets orbiting close to their host stars are expected to have smaller atmospheres (lost through evaporation) and therefore the sample used in this work may be biased toward higher densities.  Figure 1 shows histograms of the relative mass and radius error for the observed exoplanets from the NASA Exoplanet Archive \footnotemark. Currently, the distributions are peaked at $\sim 8\%$ in radius and $\sim 35\% $ in mass, but upcoming space and ground-based missions are expected to improve these values to a few percent. We used uncertainties of $3 \%, 5\%$, and $10\%$ in radius and $5 \%, 15\%$, and $30\%$ in mass for our synthetic planets to cover the range from the current most common uncertainties to the smallest uncertainties. Figure 2 shows the synthetic planets, the observed population exoplanets (from our revisited catalog in \cite{Otegi2019}), and M-R curves for idealized compositions of iron, Earth-like, and water ice.  The bulk densities of the synthetic planets cover the range from the physically motivated high limit (pure iron) to the minimum density of an observed exoplanet with a mass up to 20$M_{\oplus}$.\\

\footnotetext{exoplanetarchive.ipac.caltech.edu}

\begin{table}[h]

\centering
\caption{Mass, radius, and bulk density of the synthetic planets.}\label{tab:mrrho}
\begin{tabular}{c|ccc}
\textbf{Case}&\textbf{Mass [$M_{\oplus}$]}&\textbf{Radius [$R_{\oplus}$]}&\textbf{ Density [g/cm$^3$]}\\\hline

A&1&1&5.51\\
B&1.2&1.8&5.73\\
C&9&1.5&14.7\\
D&6&1.5&9.8\\
E&3&1.5&4.9\\
F&10&2&6.9\\
G&7&2&4.8\\
H&4&2&2.8\\
I&25&2.3&11.4\\
J&15&2.3&6.8\\
K&10&2.3&4.5\\
L&5&2.3&2.3\\
M&25&2.9&5.7\\
N&12&2.9&2.7\\
O&8&2.9&1.8\\
P&4&2.9&0.9\\
Q&15&3.5&1.93\\
R&10&3.5&1.3\\
S&7&3.5&0.9\\
T&4&3.5&0.5\\
\end{tabular}

\end{table}

\begin{figure*}[h]
\centering
  \begin{tabular}{@{}cc@{}}
    \includegraphics[scale=0.5]{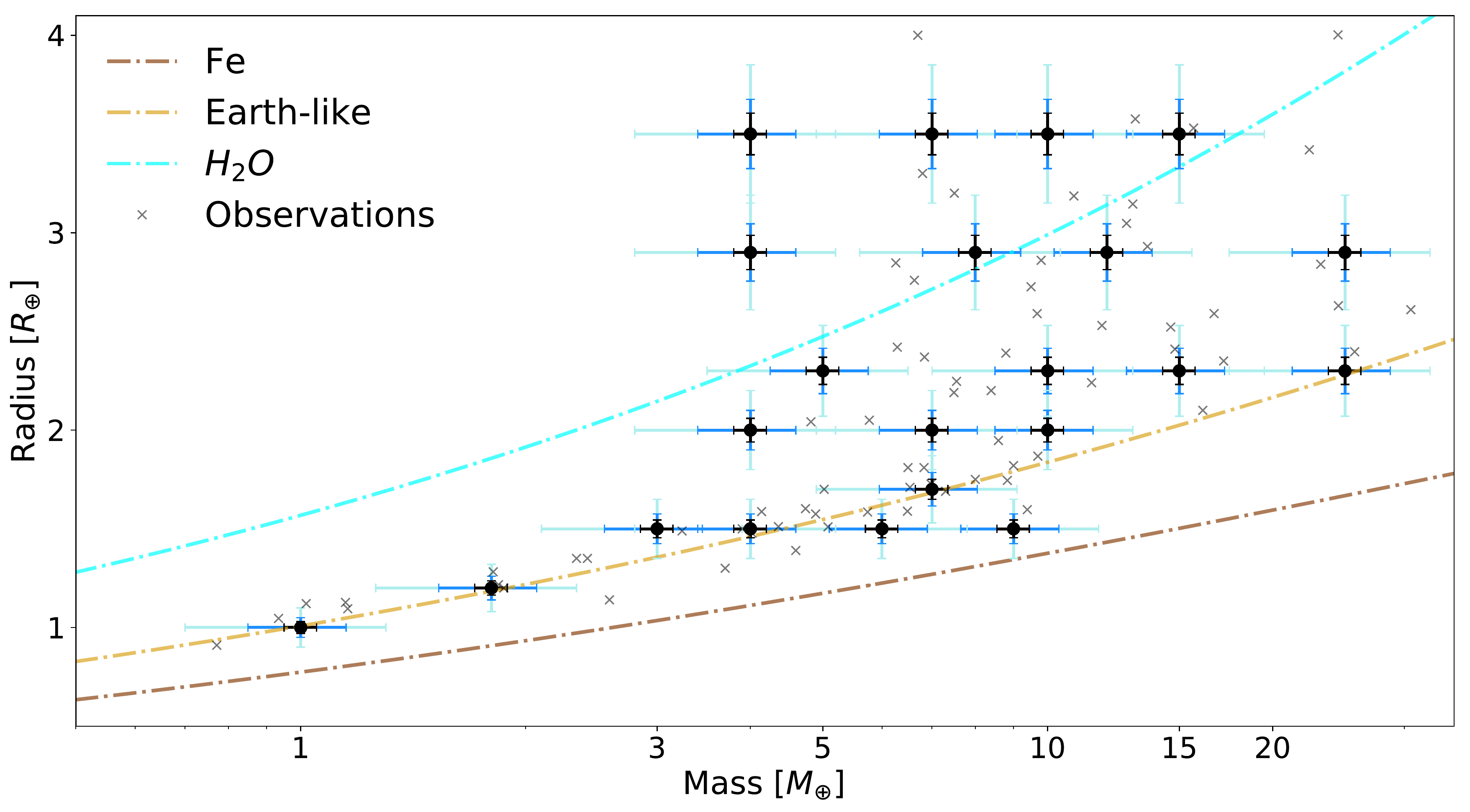}
  \end{tabular}
  \caption{M-R diagram with the synthetic planets used in the study (black dots). The black, blue, and light blue bars correspond to uncertainties of 5\%, 10\%, and 20\% in mass and 3\%, 5\%, and 10\% in radius, respectively. The crosses represent observed planets from the revisited catalog in \cite{Otegi2019}. The synthetic planets are plotted against the composition lines of pure-iron (brown), Earth-like (light brown), and water ice (blue).   }
\end{figure*}

  The irradiation has a significant effect on the interior structure determination of exoplanets with thick volatile envelopes. The exoplanets belonging to the volatile-rich population in the revisited exoplanet catalog of \cite{Otegi2019} are typically irradiated with fluxes of few hundreds times the Earth flux. We therefore used a default irradiation of $100 F_{\oplus}$ for the synthetic planets. The impact of the irradiation on the internal structure determination is studied further in section 3.5. \\

\subsection{Structure model}

We used the structure model presented in \cite{Dorn17}, which assumes a pure iron core, silicate mantle, pure water layer, and H-He atmosphere. This model uses self-consistent thermodynamics in the core, mantle, high-pressure ice, and water ocean. However in this study we used EOS for hexagonal close packed iron for super-Earth conditions presented in \cite{Hakim18} for the core density profile. These are based on density functional theory results up to 137 TPa. Unlike Earth's core \cite[e.g.,][]{Badro2007}, we did not consider the presence of light elements in the core.

The silicate mantle is assumed to be made of oxides $Na_2O-CaO-FeO-MgO-Al_2O_3-SiO_2$. Equilibrium mineralogy and density were computed as a function of pressure, temperature, and bulk composition by minimizing Gibbs free energy \citep[e.g.,][]{Connolly19}. For the water layers, we followed the approach presented in  \cite{Sotin2010}, which uses a temperate Birch-Murnaghan EOS including thermal corrections. The water can be in the solid, liquid, or super-critical phase depending on the pressure and temperature. The surface temperature of the water layer is set to be equal to the temperature of the bottom of the gas layer. \\

For the gas layer,  the equations of hydrostatic equilibrium, mass conservation, and energy transport are solved. We assumed an envelope with an elemental composition of H, He, C, and O, which are fundamental for the formation of key atmospheric molecules such as $H_2, CO, CO_2$, and $CH_4$ \citep[e.g.,][]{Madhusudhan12,Lodders02,Visscher11,Heng16}. We used the Chemical Equilibrium with Applications (CEA) package \cite[e.g.,][]{Gordon1984} for the EOS, which performs chemical equilibrium calculations for an arbitrary gaseous mixture. For the energy transport, we used the model presented in \cite{Jin14}, where an irradiated atmosphere is assumed at the top of the gaseous envelope. Within the envelope, the usual Schwarzschild criterion is used to distinguish between convective and radiative layers. More details on the structural model are found in \cite{Dorn17}. \\

\begin{figure*}[h]
\centering
  \begin{tabular}{@{}cc@{}}
    \includegraphics[scale=0.5, frame]{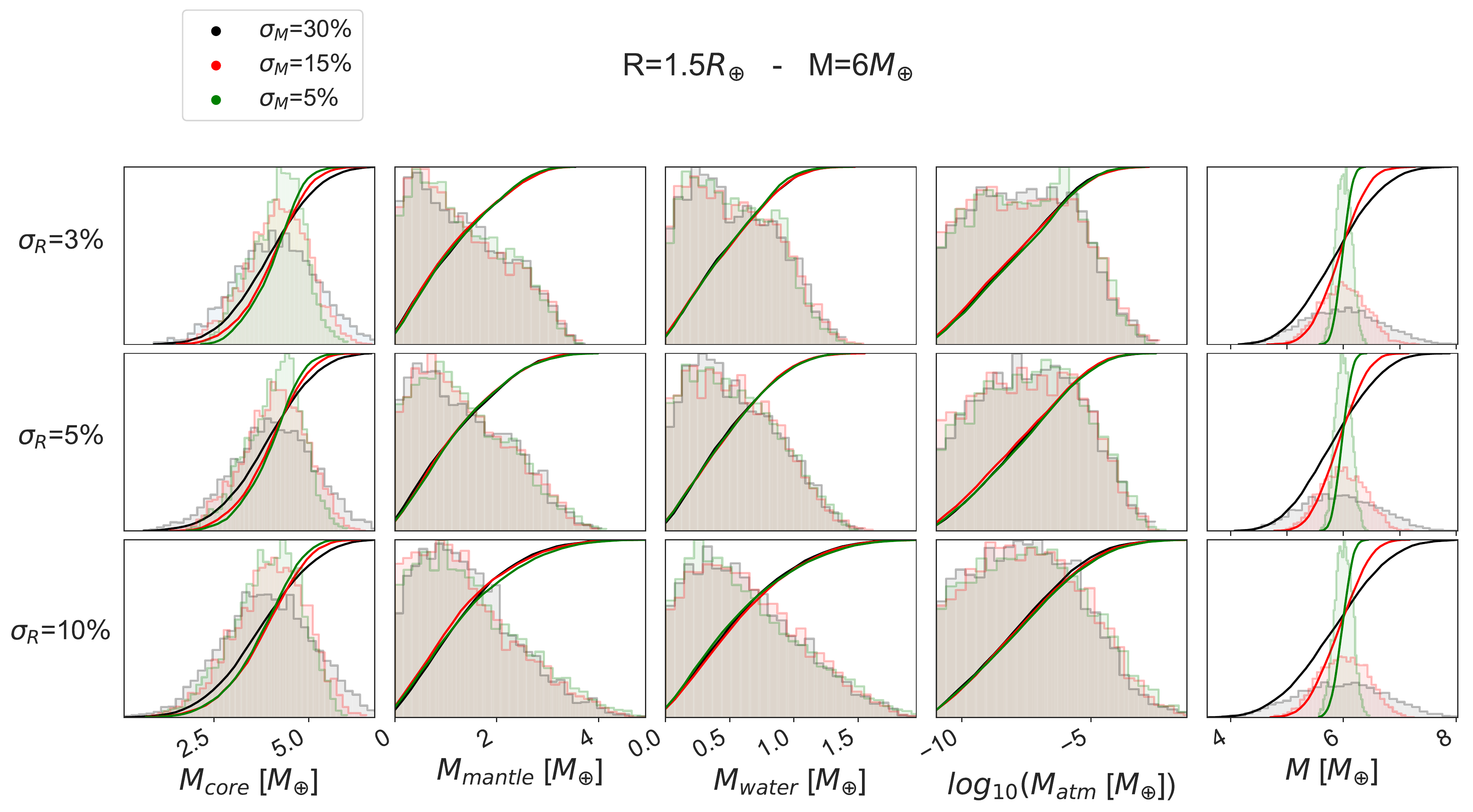}
  \end{tabular}
  \caption{Posterior distributions of core, mantle, water, atmospheric, and total masses of a planet with mass of $6 M_{\oplus}$ and radius of $1.5 R_{\oplus}$. The rows correspond to different radius uncertainties and colors in the subplots correspond to different mass uncertainties.}
\end{figure*} 

 \begin{figure}[h]
  \begin{center}
    \includegraphics[scale=0.55]{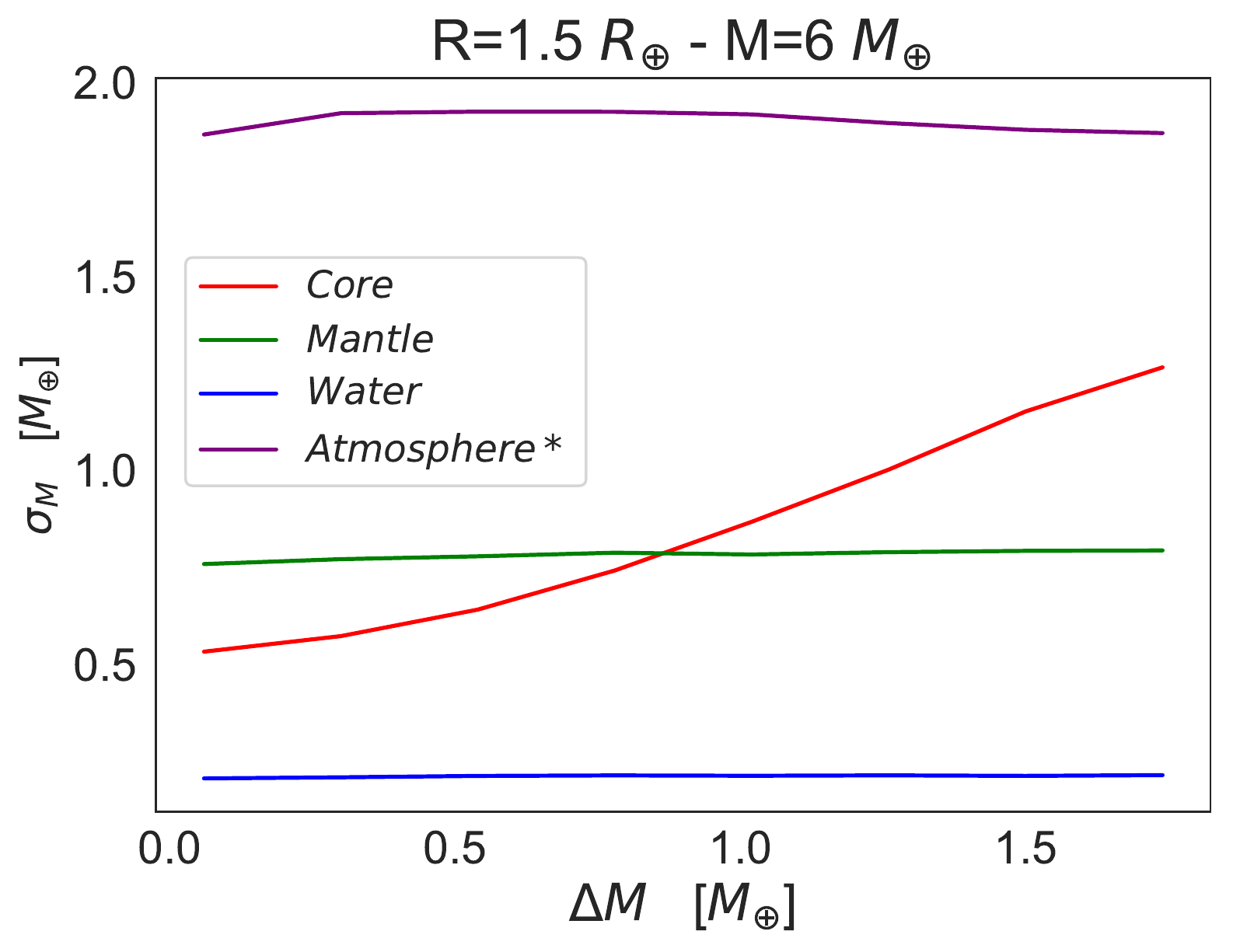}
    \caption{Standard deviation of the posterior distributions of core, mantle, water, and atmospheric masses vs.~width of the total mass distribution for a planet of $R=1.5 R_{\oplus}$ and $M=6 M_{\oplus}$. For the atmosphere, the standard deviation of the posterior distribution in logarithmic scale is used. 
    }
 
  \end{center}
\end{figure}

\subsection{Bayesian inference based on nested sampling scheme}
   
For the interior characterization we used the Bayesian inference analysis based on a nested sampling scheme using the PyMultiNest package \cite[e.g.,][]{Buchner2014}. Bayesian inference computes the posterior probability according to Bayes' theorem, which states that the probability for a fixed model parameter $\boldsymbol{x}$ given a set of data $\boldsymbol{d} $ is given by

\begin{equation}
    P( \boldsymbol{x} | \boldsymbol{d})= \frac{P( \boldsymbol{x}) \ P( \boldsymbol{d} | \boldsymbol{x})}{P( \boldsymbol{d})},
\end{equation}

where $P( \boldsymbol{x}) $ is the prior probability of $\boldsymbol{x} $ before the data is observed, $P( \boldsymbol{d}) $ the Bayesian evidence, and $P( \boldsymbol{d} | \boldsymbol{x})$ the likelihood function. The likelihood function represents the probability of observing the data given the model parameter $\boldsymbol{x}$, and is given by 

\begin{equation}
    P( \boldsymbol{d} | \boldsymbol{x}) = \frac{1}{(2 \pi )^{N/2} ( \prod _{i=1} ^N \sigma _i ^2)^{1/2}} \ exp (-\frac{1}{2} \sum _{i=1} ^N \frac{(g_i(\boldsymbol{x})-\boldsymbol{d}_i)^2}{\sigma _i ^2}),
\end{equation}

where N is the number of data points, $\sigma _i$ the uncertainties of the i$th$ datum, and g($\boldsymbol{x}$) the operator linking the model parameters with the data, that is, $\boldsymbol{d} = g(\boldsymbol{x})$. Our posterior probability distribution cannot be derived analytically, so we used a nested sampling scheme \cite[e.g.,][]{Skilling04}. The aim of Monte Carlo technique is to efficiently evaluate of the Bayesian evidence, but also to produce posterior probability distributions. The main strengths of nested sampling with respect to other sampling methods are the small amount of problem-specific tuning required and high efficiency.  \\

 \begin{figure*}[]
\centering
  \begin{tabular}{@{}cc@{}}
    \includegraphics[scale=0.53]{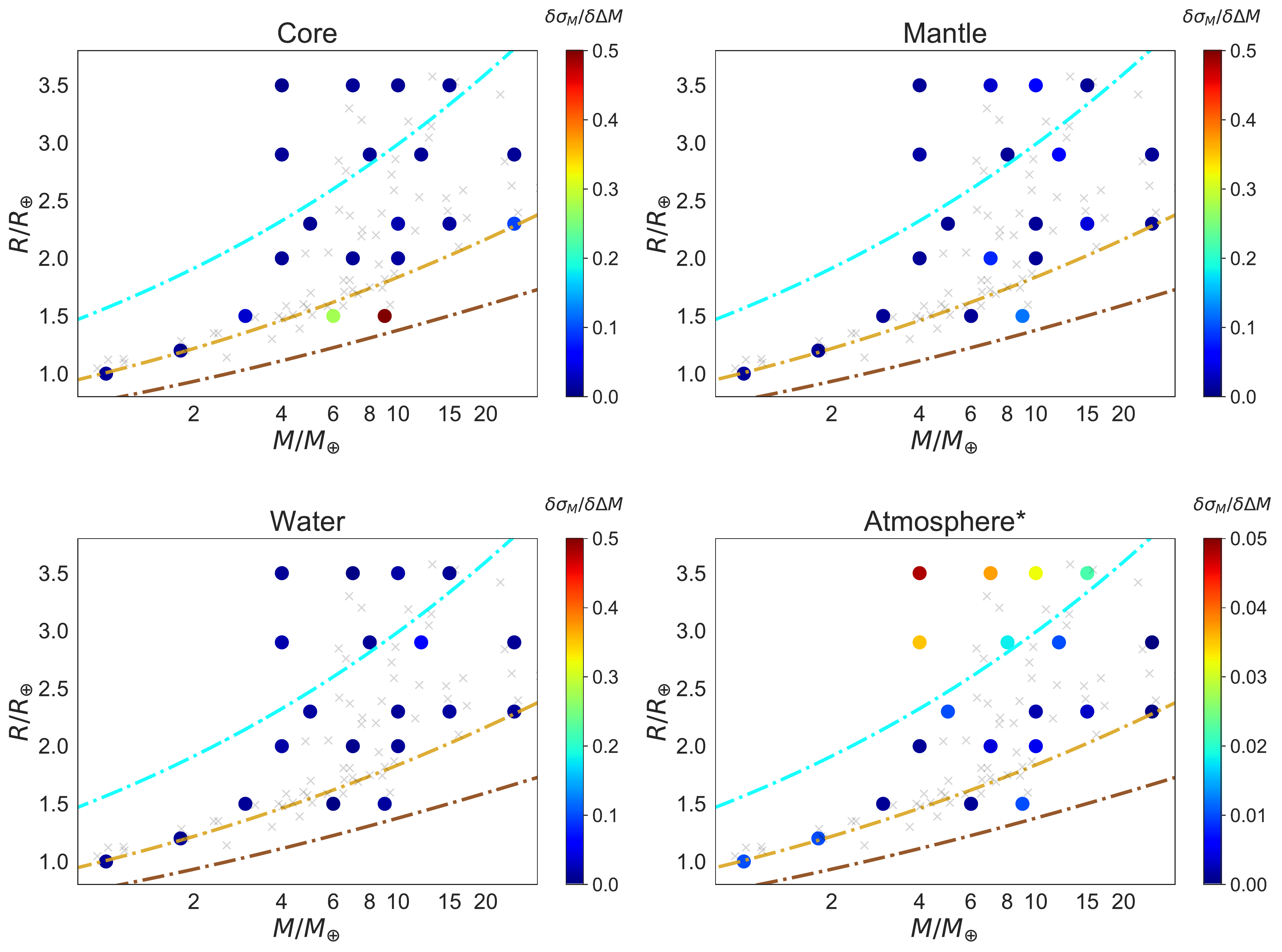}
  \end{tabular}
  \caption{  Model parameters and synthetic planets $\bm{\Delta \sigma / \Delta M} $ (see text for further details). A higher value of $\bm{\Delta \sigma / \Delta M }$ indicates that a decrease of the observational uncertainty provides additional information of the internal parameter.  The crosses represent the observed planets from Otegi et al. 2019. Composition lines of pure-iron (brown), Earth-like planet (light brown), and water ice (blue) are also plotted.} 
\end{figure*}

In short, the nested sampling scheme works as follows.
The algorithm samples some number of live points randomly from the prior $P( \boldsymbol{x} )$. The likelihood $P( \boldsymbol{d} | \boldsymbol{x})$ is evaluated at each of these points. At each iteration the point with the lowest likelihood $L_{min}$ is replaced by a new point sampled from the region of prior with likelihood $P( \boldsymbol{d} | \boldsymbol{x})> L_{min}$, keeping the number of live points constant. This process is continued until Bayesian evidence reaches some specified value (typically 0.5 in log-evidence). It produces a list of samples that can be used to produce marginalized posterior distributions. The sampling scheme is controlled by two main parameters: the number of live points and the maximum efficiency parameter. The number of live points has to be large enough to adequately sample the parameter space, for which we used a recommended value of 1000. The maximum efficiency controls the sampling volume at each iteration, which is equal to the sum of the volumes enclosing the active point set. We set it to 1 to obtain the maximum efficiency.

\subsection{Parameters and priors}

The composition of the planets are given by the masses of the iron core, mantle layer, water layer, and H-He envelope. As already mentioned we assume a pure iron core, the mantle composition is given by the Mg/Si$_\text{mantle}$, Fe/Si$_\text{mantle}$,  Al/Si$_\text{mantle}$, Ca/Si$_\text{mantle}$, and Na/Si$_\text{mantle}$ fractions. The water layer is considered to be pure H$_2$O and for the atmosphere the metallicity $Z$, luminosity $L$, and the irradiation temperature T$_\text{irr}$ are used to model the atmospheric structure. 
If not stated otherwise we fix the Mg/Si$_\text{mantle}$, Fe/Si$_\text{mantle}$,  Al/Si$_\text{mantle}$, Ca/Si$_\text{mantle}$, and Na/Si$_\text{mantle}$ fractions, $Z,$ and T$_{irr}$ while only the layer masses are varied. The luminosity is scaled as $L \propto M^{2.76}$, which fits the Jovian planets \citep[e.g.,][]{laviolette06}. 

For the nested sampling scheme we assumed a uniform prior distributions on the layer masses from 0 to the target planets mass. We would like to emphasize that for a given target planet, depending on the scientific question, a different choice of priors might be necessary. The set of prior presented here is one possible way of setting up the analysis but not the only one\footnote{For example, we could also choose to sample the layer mass fractions, instead of the layer masses, from the 3D probability simplex and use a uniform prior on the total mass of the planet.}. The impact of different priors should be subject of further studies.

As data variables we chose the total mass and total radius of the planet, except in section 3.4 where we also use the bulk Mg/Si and bulk Fe/Si fractions as data; in that case the Mg/Si$_\text{mantle}$ and Fe/Si$_\text{mantle}$ fractions are also varied, assuming a uniform prior. The total mass is calculated as the sum of all layer masses. The total radius and the bulk Mg/Si and Fe/Si fractions are an outcome of the structure model. In Table 2 we summarize the model parameters, their priors, and the data variables.

\begin{table}[h]
\centering
\caption{Summary of model parameters, priors, and data.}
\begin{tabular}{llll}
\hline
\textbf{Model Param.} &&\textbf{Priors}& \textbf{Data}\\
\hline
$M_{Core}$ &&$\mathcal{U}(0,M)$& Total Mass\\
$M_{Mantle}$ &&$\mathcal{U}(0,M)$& Total Radius\\
$M_{Water}$ &&$\mathcal{U}(0,M)$& Bulk Fe/Si\\
$M_{H-He}$ &&$\mathcal{U}(0,M)$& Bulk Mg/Si\\
$Fe/Si_{Mantle}$ &&const. or $\mathcal{U}(0,2  Fe/Si_{\odot})$& \\
$Mg/Si_{Mantle}$ &&const. or $\mathcal{U}(0,2  Mg/Si_{\odot})$&\\
$Al/Si_{Mantle}$ &&const.&\\
$Ca/Si_{Mantle}$ &&const.&\\
$Na/Si_{Mantle}$ &&const.&\\
Z &&const.&\\
L &&const.&\\
$T_{irr}$ &&const.&\\
\hline
\end{tabular}
\end{table}

   
\section{Dependence of internal structure determination on observational uncertainties}
In this section we explore how the internal structure determination depends on the observational uncertainties for exoplanets of different masses and radii fixing the irradiation to 100$F/F_{\oplus}$ and solar Fe/Si and Mg/Si abundances. In particular, we explore how the posterior distribution of the internal parameters depend on the observational uncertainty for the synthetic planets in Table 1. An example is shown in Figure 3, in which the posterior distributions of the internal parameters corresponding to a planet of observed mass and radius of $6 M_{\oplus}$ and $1.5 R_{\oplus}$ with different observational uncertainties are shown. The first four columns show the posterior distributions of the iron core, mantle, water, and atmospheric masses (output) renormalized by the sum, and the last column shows the mass distribution of the planet (input). The rows correspond to different observed radius uncertainties and the colors to different observed mass uncertainties; black, red, and green for 30\%, 15\%, and 5\%, respectively.  We find that for a planet of $6 M_{\oplus}$ and $1.5 R_{\oplus}$ a decrease in the uncertainties improves the determination of the core mass, but has a negligible effect on the determination of the other internal parameters.\\

\begin{figure}[h]
  \begin{center}
    \includegraphics[scale=0.29]{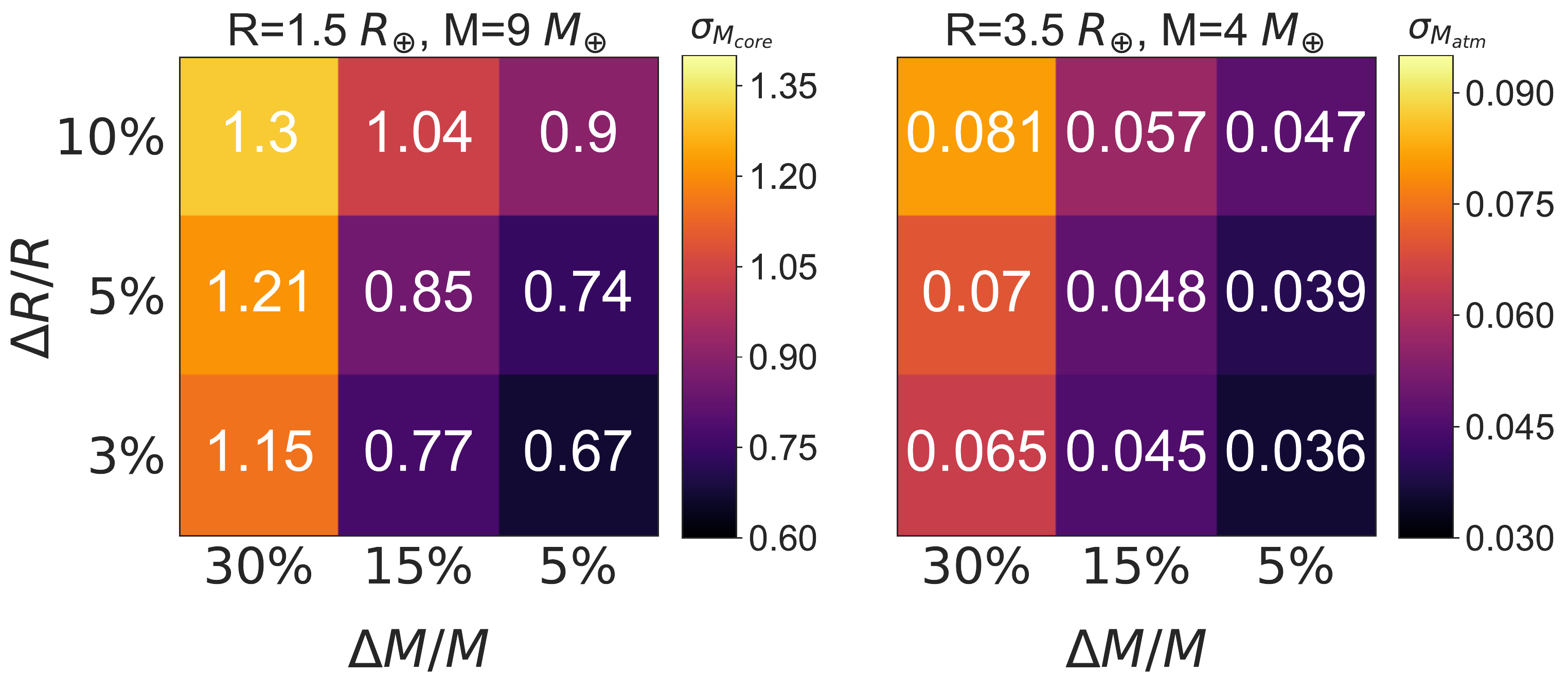}
    \caption{ Observational uncertainties in mass and radius versus $\bm{\sigma_{core}}$ (left) and $\bm{\sigma_{atm}}$ (right) for the two synthetic planets with the highest and lowest densities.}
 
  \end{center}
\end{figure}

In Figure 4 the standard deviation ($\sigma$) of the posterior distributions of the internal structure parameters are shown versus the mass uncertainty ($\Delta M$), for a radius of 1.5$R_{\oplus}$ and a mass of 6$M_{\oplus}$. We aim to use this approach to explore what interior parameters are better constrained when the mass precision is higher, keeping the radius fixed with zero uncertainty. In this work we assume a flat uniform distributed uncertainty for the planetary mass since it simplifies the analysis and allows us to consider cases with uncertainties close to 0. The equation $\Delta M=0$ shows the  intrinsic degeneracy of the interior parameters when data uncertainty is zero for a planet of $1.5 R_{\oplus}$ and $6 M_{\oplus}$. When $\Delta M$ increases, $\bm{\sigma}$ of the core mass increases, meaning that this is the only tested interior parameter that is sensitive to a change in observational uncertainties. It is in agreement with what is shown in Figure 3 where we use a more realistic Gaussian distribution for the mass and radius.  It is interesting to note that the evolution of $\sigma _M$ with $\Delta M$ is nearly linear. The slope $\bm{\delta  \sigma _M / \delta \Delta M}$ contains very valuable information: it indicates how much  interior estimates (i.e., layer mass fractions) can be improved by increasing data precision (i.e., mass). 
Figure 5 shows these slopes  $\delta  \sigma _M / \delta \Delta M$ of the core, mantle, water, and atmospheric masses for all synthetic planets. The results allow us to differentiate three different regimes, described in the following subsection.

\subsection{Planets below Earth-like composition line}

 We refer to planets C and D (Table \ref{tab:mrrho}). For these planets, a decrease of the observational uncertainties leads to a better determination of the core mass.  In this density regime, planets do not have a significant gaseous envelope and they require a large amount of iron. Then, variation in the mass or radius distributions is reflected in the core mass, since there is no other interior layer that can account for such high planet bulk densities. \\

As mentioned above, Figure 5 is constructed assuming flat uniform error distributions for mass and radius. To check whether the conclusions are also valid for more realistic distributions we compare the results using the Gaussian and flat distributions, where $\sigma$ of the Gaussian is equal to $\Delta M$ of the flat distribution. The obtained results with the flat distributions are consistently slightly higher than with the Gaussian results, but the general shape is nearly equal. Figure 6 shows $\bm{\sigma_{core}}$ versus observational uncertainties in mass and radius using Gaussian distributions. It shows that $\bm{\sigma}$ does not decrease uniformly with uncertainty and that the internal structure determination is more sensitive to an improvement of data uncertainties when data uncertainties are large.  Furthermore, we find that a decrease of the observed mass uncertainty is much more effective to better determine the core mass than a decrease in the radius uncertainty. This is because the core mass directly affects the planetary mass, while its effect on radius is smaller. \\

\subsection{Planets above pure-water composition line}

\begin{figure*}[h]
\centering
  \begin{tabular}{@{}cc@{}}
    \includegraphics[width=\linewidth]{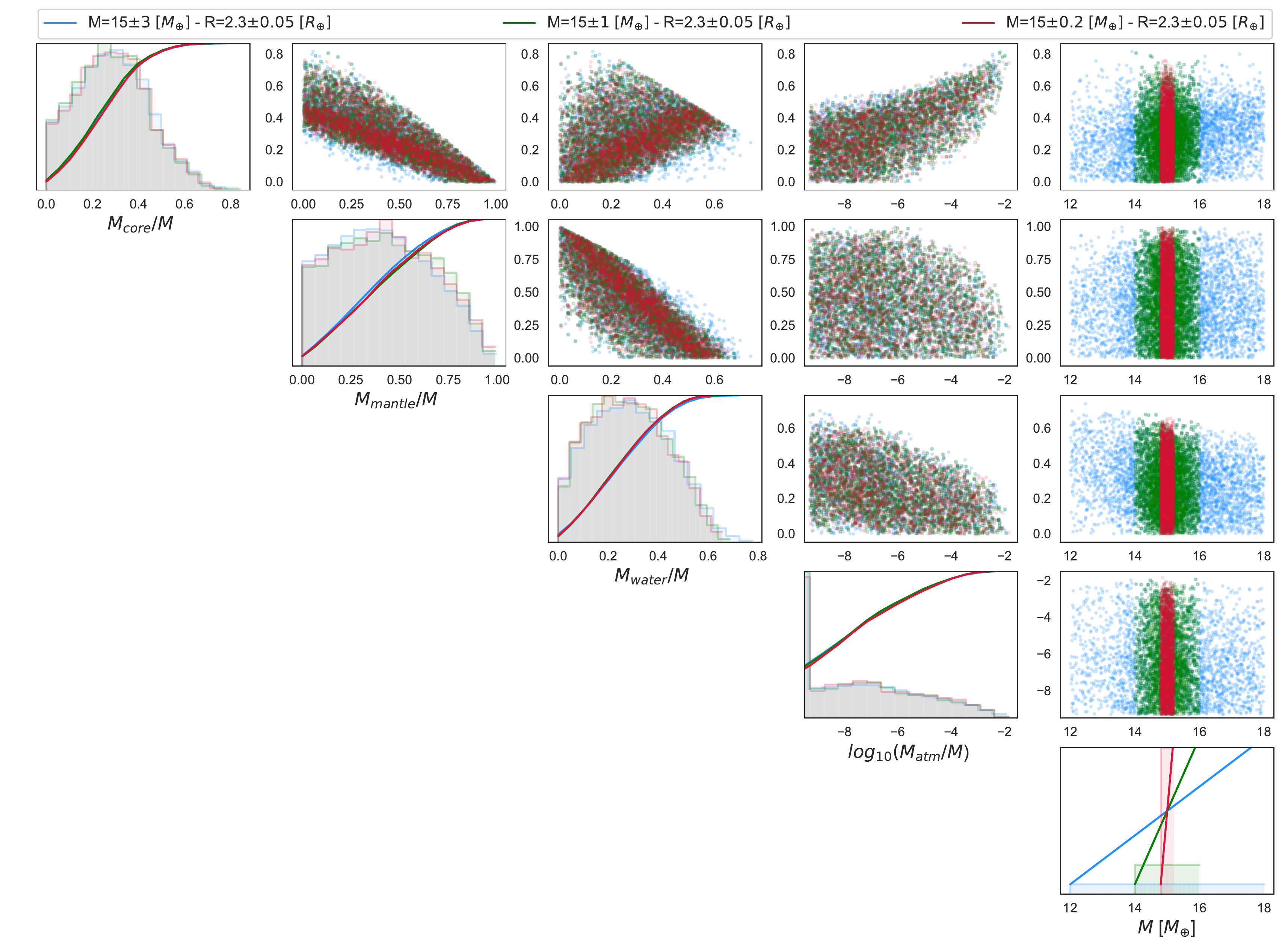}
  \end{tabular}
  \caption{Illustration of the degeneracy in the intermediate density regime previously defined. The posterior distributions the internal parameters of a planet with fixed radius of $2.3 R_{\oplus}$ and mass of $15 M_{\oplus}$ are shown The different colors represent different widths of the mass distribution.}
\end{figure*}

We find that when a planet lies above the pure-water composition line a decrease in the mass and radius uncertainties leads to a better determination of the H-He envelope mass. The explanation is similar to that discussed above for high-density planets: for a given mass, low-density planets require thick gaseous envelopes to match the observed radius, and this significantly reduces the degeneracy. Therefore a change in the total mass/radius distributions has a direct impact on the properties of the gaseous envelope (i.e., mass fraction, thickness).  

The right panel of Figure 6 shows the $\bm{\sigma _{atm}}$ versus the observational uncertainties in mass and radius for the low-density synthetic planets with realistic Gaussian mass and radius distributions. Interestingly, in this regime  decreasing the radius uncertainty is more effective than decreasing the mass uncertainty to constrain the atmospheric mass. 
This can be understood by the fact that the atmospheric mass fraction of the low-density planets in the synthetic sample is nearly negligible, but the radius fraction is significant. \\

\subsection{Planets between Earth-like and pure-water composition lines}

Finally, we find that planets with densities between the Earth-like and pure-water composition lines are the most degenerate. For these planets, decreasing the mass and radius uncertainties does not improve the determination of any internal structure parameter. Even when removing one compositional layer, there is a substantial degeneracy among the other three. Figure 7 shows the posterior distributions the interior parameters of a planet with fixed radius of $2.3 R_{\oplus}$ and mass of $15 M_{\oplus}$. The different colors represent different widths of the total mass distribution. The posterior distributions of the internal parameters do not cover the same region in the two-dimensional parameter space. The posterior distributions only overlap when projecting them into a one-dimensional histogram. Therefore, for these planets it would be crucial to have additional information that can further constrain the internal structure,  such as as atmospheric metallicity measurements from space missions like ARIEL (Atmospheric Remote-sensing Infrared Exoplanet Large-survey; e.g., \cite{Tinetti2016}) and JWST (James Webb Space Telescope; e.g., \cite{Beichman18}).   \\

\subsection{Using stellar abundances as an additional constraint}

\begin{figure*}[h]
\centering
  \begin{tabular}{@{}cc@{}}
    \includegraphics[scale=0.48]{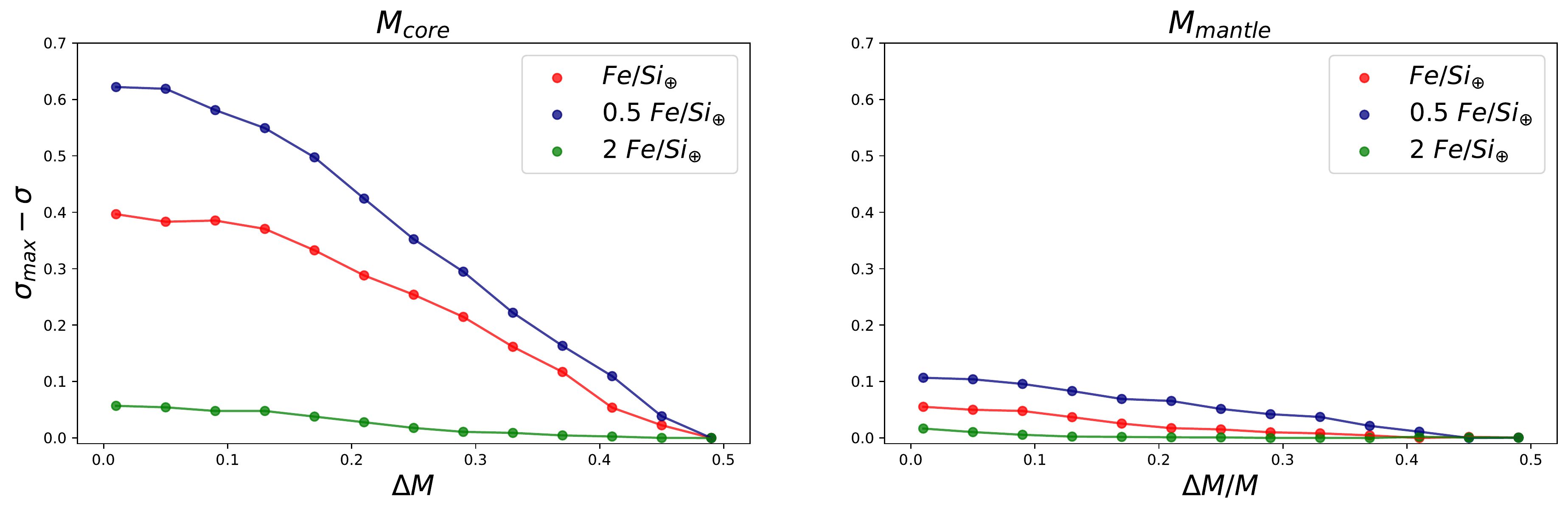}
  \end{tabular}
  \caption{Standard deviation of the core mass (left) and mantle mass (right) for a planet of mass $15 M_{\oplus}$ and radius $2.3 R_{\oplus}$ vs. the width of the mass distribution. The red, green, and blue lines represent different Fe/Si in the planet. A higher value of $\sigma_{max}-\sigma$ indicates that the internal parameter is better constrained with respect to the reference, at which $\sigma=\sigma_{max}$.}
\end{figure*} 

\begin{figure*}[h]
\centering
  \begin{tabular}{@{}cc@{}}
    \includegraphics[scale=0.48]{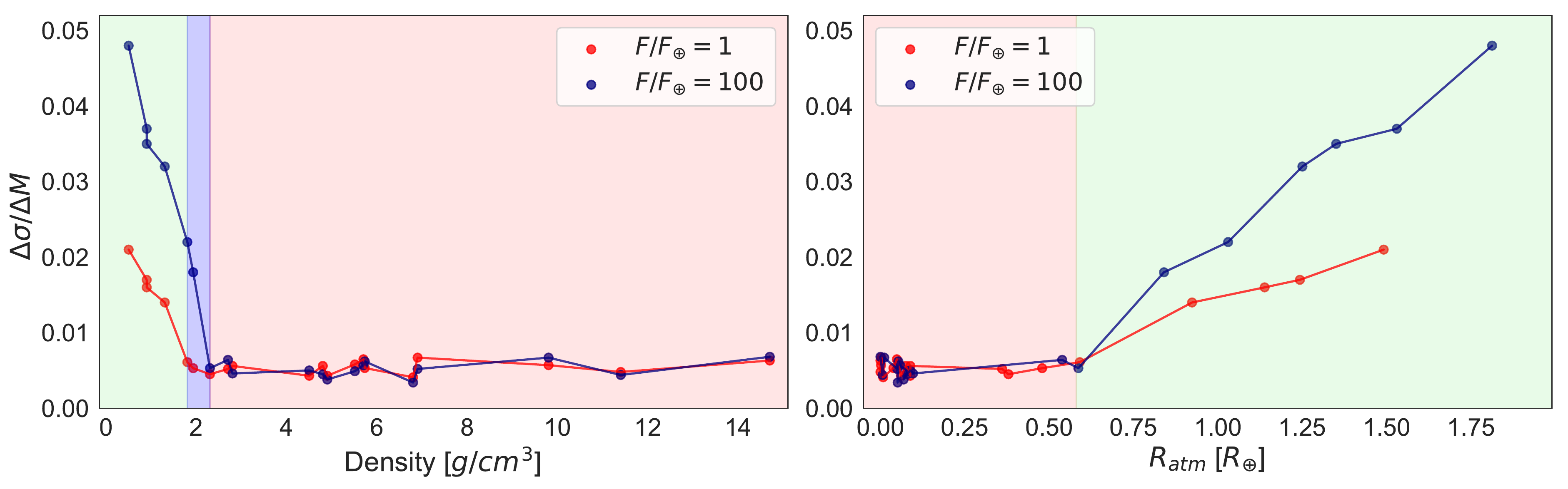}
  \end{tabular}
  \caption{ Atmospheric mass $\bm{\Delta \sigma / \Delta M }$ for the synthetic planets vs. the bulk density (left) and median atmospheric thickness (right) for different irradiation. The light green and red regions represent the ranges in which decreasing the observational uncertainties lead and do not lead to better constraints on the atmospheric mass, respectively. The blue region represents the density range in which planets with an irradiation of 100 Earth fluxes have better constrained atmospheric mass with decreasing uncertainties and planets with an irradiation of one Earth flux do not.  }
\end{figure*}

As shown in \cite{Dorn17}, abundance constraints of Fe/Si and Mg/Si from the host star can serve as a proxy for the planet bulk abundance and can reduce model degeneracy. This assumption reproduces the composition of the Earth \citep[e.g.,][]{Javoy10} and Mars \citep[e.g.,][]{Khan08}, but not the one of Mercury, for which a post-formation giant impact scenario is often considered \citep[e.g.,][]{Benz07,Chau18}. \cite{Thiabaud15} used a chemical model to link the composition of  18 synthetic stars with solar mass and luminosity with the composition of the hosted planets, and they found a close relation. Nevertheless, \cite{Wang18} showed that there are some differences for the Earth compared to the Sun, and that volatilization trends in the bulk composition of exoplanets should be considered \cite[e.g.,][]{Dorn2019}. \\

We next explore how the results are affected when assuming that the stellar relative abundances reflect the planetary relative abundances. 
We then use the bulk Fe/Si and Mg/Si as additional constraints, and introduce two additional model parameters: the Fe/Si and Mg/Si ratios in the mantle. The goal of this section is to evaluate whether using the bulk abundances allows us to improve the estimation of the internal structure using a more detailed interior model. 
We find that using the bulk Fe/Si and Mg/Si does not necessarily further constrain the planetary internal structure when adding Fe/Si and Mg/Si ratios in the mantle as free parameters. We assume a small uncertainty on the bulk Fe/Si and Mg/Si of 3\%. 
Figure 8 shows the standard deviation of the core mass and mantle mass for a planet of the intermediate-density regime in terms of the width of the total mass distribution. In this case $\bm{\sigma_{max}-\sigma}$ is displayed, so that a higher value means that the interior parameter is better constrained with respect to the reference, at which $\bm{\sigma=\sigma_{max}}$. In the case of the chosen synthetic planets, when we introduce the constraint of $\text{Fe/Si}_\text{planet} = \text{Fe/Si}_\text{star}$ with $\text{Fe/Si}_\text{star}<\text{Fe/Si}_{\odot}$, their core mass fraction is better constrained. The lower $\text{Fe/Si}_\text{star}$, the better the core mass gets constrained. Regarding their mantle mass fraction, it is only better constrained when $\text{Fe/Si}_\text{star}< 1/2 \ \text{Fe/Si}_{\odot}$. 
Considering the bulk stellar abundances mostly affects the core mass fraction determination. For $\text{Fe/Si}_\text{star}> 2 \ \text{Fe/Si}_{\odot}$ the abundance constraints do not provide additional information on the internal parameters for the tested planets. 
\\

\subsection{Sensitivity of atmosphere-rich planets to irradiation}

 \begin{figure*}[h]
\centering
  \begin{tabular}{@{}cc@{}}
    \includegraphics[scale=0.5]{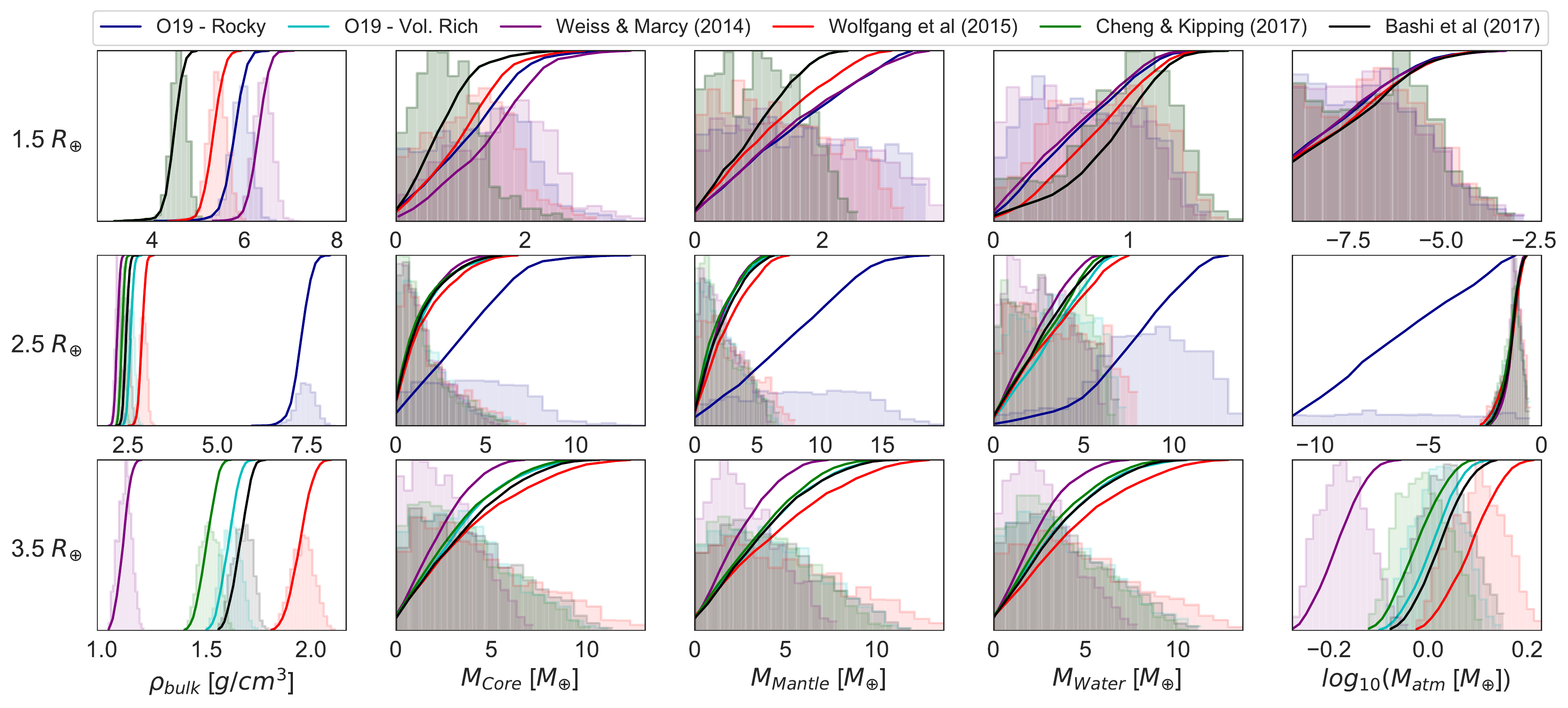}
  \end{tabular}
  \caption{Posterior distributions of the bulk density and internal parameters of the planet for different radii and masses computed with various published M-R relationships. }
\end{figure*}

\begin{table*}[h]
\centering
\caption{Masses from published M-R relationships for different radii. O19-R: \cite{Otegi2019}--rocky population; O19-V: \cite{Otegi2019}--volatile-rich population; W14: \cite{Weiss14}; W15: \cite{Wolfgang15}; C17: \cite{Chen17}; B17: \cite{Bashi17}. }
\begin{tabularx}{\textwidth}{X||XX|XXXX}

&\textbf{O19-R}&\textbf{O19-V}&\textbf{W14}&\textbf{W16}&\textbf{C17}&\textbf{B17}\\\hline

\textbf{1.5 $R_{\oplus}$}& 3.64 $M_{\oplus}$& - &3.92 $M_{\oplus}$&3.32 $M_{\oplus}$&2.80 $M_{\oplus}$& 2.79 $M_{\oplus}$\\
\textbf{2.5 $R_{\oplus}$}& 21.24 $M_{\oplus}$& 7.4 $M_{\oplus}$&6.30 $M_{\oplus}$&8.32 $M_{\oplus}$&6.66 $M_{\oplus}$& 7.05 $M_{\oplus}$\\
\textbf{3.5 $R_{\oplus}$}& -& 12.59 $M_{\oplus}$&8.62 $M_{\oplus}$&15.25 $M_{\oplus}$&11.77 $M_{\oplus}$& 13.00 $M_{\oplus}$\\

\end{tabularx}

\end{table*}

The calculations presented above correspond to an irradiation of $100 F_{\oplus}$. 
In this subsection we explore the sensitivity of the results to the assumed stellar irradiation. 
It should be noted that our model does not include atmospheric evaporation  and the effect of the insolation is only reflected on the equilibrium temperature of the planet. Figure 9 shows $\bm{\Delta \sigma / \Delta M }$ of the envelope mass for all the simulated planets versus~the bulk density and atmosphere radius for different assumed insolations. 
Clearly, there is a difference when changing the insolation for planets with relatively large atmospheres. A higher insolation increases the temperature of the H-He atmosphere, which leads to an expansion and a decrease in H-He layer density. In general, the atmospheric mass can be better constrained for strongly irradiated planets.

Figure 9 shows whether an improvement in observational uncertainties contributes to constrain the envelope mass depending on the planetary density and irradiation.  Planet O ($M=8 M_{\oplus},R=2.9 R_{\oplus}$) has a null $\bm{ \Delta \sigma / \Delta M}$ for Earth-like irradiation, but it is positive for high irradiation. If we look at this depending on the atmosphere radius (median of the posterior distribution) instead of the density, we see that the breakpoint coincides for the two insolations.  The determination of atmospheric mass of planets holding atmospheres up to $0.5 R_{atm}$ is insensitive to an improvement of observational uncertainties. When the radius of the H-He envelope is lower than $0.5 R_{atm}$ there is a strong degeneracy with the other layers, but otherwise a high amount H-He is needed to fit the observed mass and radius, and therefore,  variations on the observational uncertainties affect the determination of the H-He envelope. It is important to note that a more realistic model would include water in the atmosphere. However, this would be another free parameter that would increase the degeneracy and would be difficult to determine.

\section{Dependence of internal structure determination on the mass derived from M-R relationships. }

Since not all planets have measured masses and radii, (e.g., most Kepler planets in the past), published M-R relationships are often used to estimate the mass of a planet for a given radius and vice versa. Figure 4 of \cite{Otegi2019} compares some of the published mass radius relations. This figure shows significant disagreement in mass of approximately $25 \%$ for a given radius. Despite these differences, given the degeneracy when determining the internal structure, different masses inferred from various M-R relations could lead to very similar internal structures. In such a case, even if only the radius is measured, it is still possible to infer information on the planetary structure and bulk composition. In this section we explore whether these differences in mass are reflected in the determination of the internal structure parameters, or whether they get diluted because of the degeneracy. \\

 Figure 10 shows the posterior distribution of the internal structure parameters for planets with the same radii and different masses as computed from the published M-R relationships (the masses and radii are listed in Table 3). We do not include other published M-R relationships as those presented in \cite{Weiss13} or \cite{Zeng16} because they also depend on an additional parameter (irradiation and core mass fraction, respectively). Since for a radius of $1.5 R_{\oplus}$ the masses corresponding to \cite{Chen17} and \cite{Bashi17} are almost identical we only show one of them (\cite{Chen17}). We assume a very small uncertainty of $1\%$ for mass and radius. \\ 

\begin{figure*}[h]
\centering
  \begin{tabular}{@{}cc@{}}
    \includegraphics[scale=0.43]{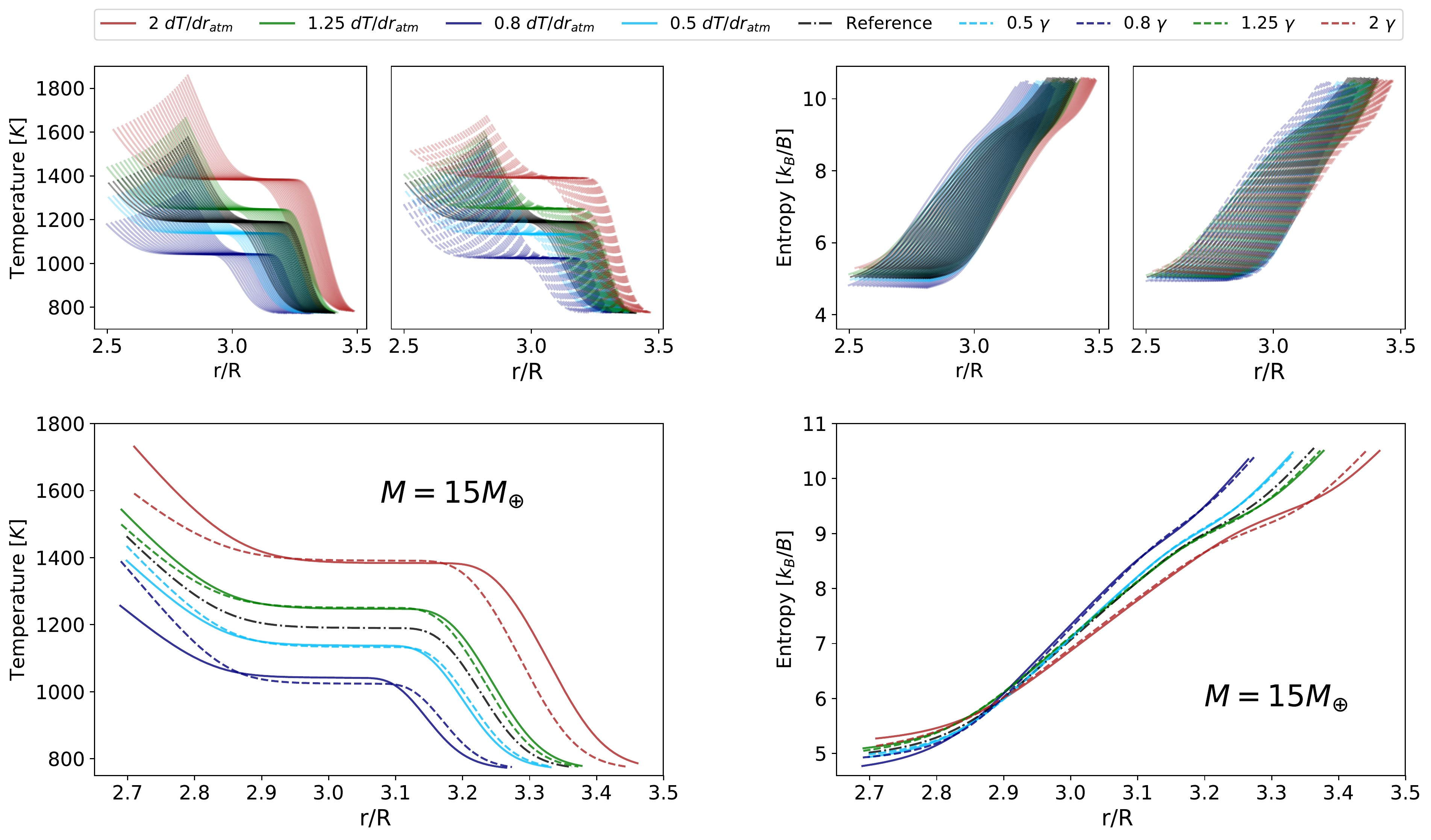}
  \end{tabular}
  \caption{Atmospheric temperature and entropy profiles when artificially shifting the temperature gradient (solid lines) and $\gamma$ (dashed lines). The temperature and entropy profiles of 25 planets with masses from $3 M_{\oplus}$ to $20 M_{\oplus}$ (from bottom to top) with a fixed composition of $0.5\%$ H-He mass fraction, $40\%$ water mass fraction, and $59.5\%$ of a rocky core. On the bottom, the profiles corresponding to a mass of $15 M_{\oplus}$.  }
\end{figure*}

\begin{figure*}[h]
\centering
  \begin{tabular}{@{}cc@{}}
    \includegraphics[scale=0.52]{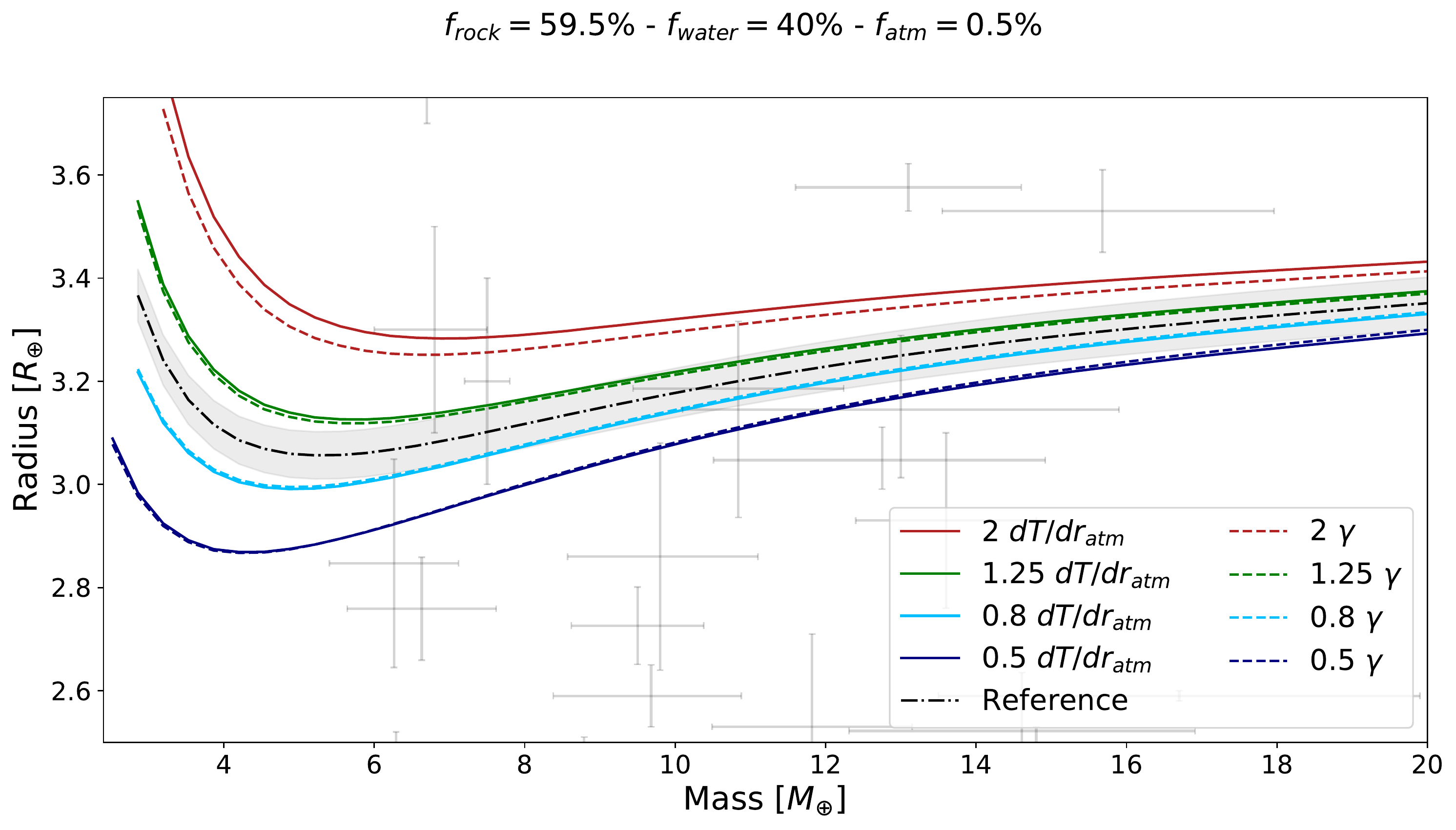}
  \end{tabular}
  \caption{M-R relationships for a planet with fixed composition and the perturbations applied to the temperature profile as presented in Figure 11. The gray envelope corresponds to a radius uncertainty of $3 \%$, which will soon be reached with future missions. }
\end{figure*}

For a radius of $1.5R_{\oplus}$ the masses from different M-R relationships lead to significantly different inferred core mass distributions. The mass from \cite{Weiss14} leads to an inferred core mass of $2.1 \pm 0.7 M_{\oplus}$ and \cite{Chen17} to $0.8 \pm 0.5M_{\oplus}$. At this radius the bulk densities lie between 4g/cm$^3$ and 7g/cm$^3$, so these planets are expected to have a negligible gaseous envelope. In addition, the water mass is small compared to the mass of refractory materials. Consequently, the degeneracy is reduced and the internal structure is more sensitive to a change in bulk density. As previously discussed, a change in the bulk density of planets dominated by refractory materials is mainly reflected in the core mass, and for this reason the core mass distributions for planets with $1.5R_{\oplus}$ are sensitive to slight mass variations. There are some minor differences between the mantle and water mass distributions, especially between \cite{Chen17} and the other considered studies.\\

For a radius of $2.5 R_{\oplus}$ the mass computed from the M-R relationship for rocky exoplanets in \cite{Otegi2019} is much higher compared to the masses from other published M-R relationships, and consequently the inferred internal structure is drastically different. Nevertheless, the difference between the internal structures inferred with the masses from other M-R relationships are very similar to each other. They lie in the intermediate density regime previously discussed, which is mostly degenerate. Consequently, there are not significant differences in the internal structure independently of the M-R relationship used, except for that corresponding to rocky exoplanets as in \cite{Otegi2019}; this work estimates a mass for $2.5 R_{\oplus}$ which is approximately three times larger than the other M-R relationships. \\

Finally, for radius of $3.5 R_{\oplus}$  \cite{Weiss14} leads to much lower masses than the other M-R relationships, and therefore it leads to a significantly different estimated internal structure. The main difference at this radius arises in the posterior distribution of the atmospheric mass. We find that the choice of the M-R relationship for internal structure determination only in the solid-dominated and volatile-rich radius regimes is relevant. \\ 

We conclude that planetary internal structure should not be inferred from the radius alone. Even if for radii smaller than 2$R_{\oplus}$ most of the observed exoplanets closely follow the M-R relationships, relatively small variations in mass have a significant impact on the inferred core mass fraction. For radii in the range 2-3$R_{\oplus}$ where the degeneracy is the strongest, instead, small variations on the mass lead to almost identical inferred internal structure. However, this radius regime corresponds to the transition from rocky to volatile-rich exoplanets, and the large diversity of masses for a given radius does not allow us to infer the internal structure by radius alone. 

\section{Dependence of the radius on the temperature profile.}

Internal structure models depend on the assumed EOS and on several theoretical assumptions such as the internal structure (mixed, differentiated), the envelope structure and its luminosity, and the composition of the planetary layers. Since observational uncertainties are expected to decrease  significantly in the upcoming years, it is desirable to explore whether the theoretical uncertainties related to the model assumptions could dominate compared to the observational uncertainties. 
In this section we investigate how the assumed temperature profiles affect the inferred planetary radius and study what uncertainty in the temperature profile of a planet is required to match the observational uncertainty. 

Relative changes in material density have the largest effects on the radius for low-density materials, for example, variations in the metallicity distribution in the envelope significantly affect the inferred planetary radius \cite[e.g.,][]{Lozovsky18}. 
We therefore explore how variations in the density, temperature, and entropy profiles in the H-He envelope affect the radius of a planet.
A common assumption in the envelope structure models is to consider a fully adiabatic temperature profile. However, more realistic models should account for compositional gradients along the envelope, which leads to non-adiabatic temperature profiles. \textbf{We do not know a priori what composition gradients might exist in most planets}. In this section, we artificially perturb the temperature profile from the adiabat and study how it affects the inferred planetary radius. \\
We perturb the temperature profiles in two different ways: shifting the temperature gradient $dT/dR$ in the atmosphere and the $\gamma$ as defined in \cite{Hansen08}, which is the ratio of visible to infrared opacities. The perturbations in $dT/dR$ and $ \gamma$ are selected to produce a change in the radius comparable to the observational uncertainties. The effect of these perturbations on the temperature and entropy profiles is shown in Figure 11.  The top panels show the temperature and entropy profiles of 25 planets with masses from $3 M_{\oplus}$ to $20 M_{\oplus}$ and different assumed atmospheric temperature gradients and $ \gamma$. The bottom panels illustrate the effect of these variations in the internal temperature and entropy profiles of a planet with a mass of  $15 M_{\oplus}$. Figure 12 shows the composition lines corresponding to a fixed composition and the perturbations applied to the temperature profile as presented in Figure 11. The gray envelope corresponds to a radius uncertainty of 3\%, which is expected to be reached with future missions. We find that small variations in the temperature profile, such as those corresponding to uncertainties of 20\% in the temperature gradient or the ratio of visible and infrared opacities lead to radius uncertainties of 3\%. 
This implies that theoretical uncertainties that are associated with the model assumptions could be larger than the observational uncertainties and therefore dominate the uncertainties in internal structure models. It would be necessary to do a systematic and detailed analysis of the uncertainties introduced by the model assumptions (e.g., EOS, composition of the layers, layer boundaries, intrinsic luminosity of the planet, and atmospheric opacities).

\section{Summary and conclusions}

We present new internal structure models based on the work of  \cite{Dorn17} with a Bayesian inference analysis using a nested sampling scheme. We explore several aspects that affect the characterization of exoplanets with masses up to 25$M_{\oplus}$ and radii up to 3.5$R_{\oplus}$, such as how variations in the observational uncertainties or location in the M-R affect the inferred internal structure, how the choice of the mass using different M-R relationships for a fixed radius influences the inferred internal structure, and how the atmospheric temperature profile affects the inferred radius. We should keep in mind that these results were computed for a given set of priors. The sensitivity of the results given various priors should be a topic for future investigation. Regarding the sensitivity of internal characterization to observed mass and radii, our main findings are summarized as follows: 

\begin{itemize}
    \item A decrease in observational uncertainties for planets below the Earth-like composition line leads to a better determination of the core mass. 
    
    \item A decrease in observational uncertainties for planets above the pure-water composition line leads to a better determination of the atmospheric mass.
    
    \item A decrease in observational uncertainties for planets between Earth-like and pure-water composition lines does not significantly improve the determination of any internal structure parameter. 
    
    \item The density boundaries listed above slightly depend on the used interior model (e.g., luminosity, insolation). 
    
    \item The atmospheric mass of strongly irradiated atmosphere-rich planets can be better determined than for weakly irradiated planets. 

    \item Using the stellar Fe/Si and Mg/Si abundances as a proxy for the bulk planetary abundances does not always help to constrain the planetary internal parameters when adding two extra model parameters related to the mantle composition. This depends on the actual value of the measured stellar abundances and their uncertainties and on the data of planetary mass and radius. Low stellar Fe/Si ratios improve the determination of the core mass for planets with densities between 2.3g/cm$^3$ and 6.9g/cm$^3$, where the planetary internal parameters are most degenerate. 
    
    \end{itemize}

    We also find that internal structure must not be estimated using radius alone. For a fixed radius, the inferred planetary internal structure using different masses from various published M-R relationships can vary significantly depending on the measured radius. For radii of nearly $1.5 R_{\oplus}$ the choice of the M-R relationship significantly affects the inferred core mass, and for radii above $3 R_{\oplus}$ this choice significantly affects the inferred atmospheric mass. For planets with radii of nearly $2.5 R_{\oplus}$ small differences in mass lead to similar inferred internal structures, but the wide range of possible masses does not allow us to infer the internal structure with radius alone.\\
    
    We find that uncertainties of $20 \% $ in the temperature gradient or the ratio of visible and infrared opacities $\gamma $ can lead to radius uncertainties of $3 \%$. Observational uncertainties are expected to decrease significantly in the near future, thereby decreasing the uncertainties in both mass (with advanced spectrographs like ESPRESSO) and radius (with next generation space missions like TESS, CHEOPS or PLATO). 
While these improved data will certainly help us to better understand planetary populations around other stars, a detailed characterization of individual planets is expected to remain somewhat limited. 
While the characterization of volatile-rich and dense exoplanets is expected to improve with decreasing observational uncertainties, a significant degeneracy in the internal structure of  most of the super-Earth population is expected to remain. 
We also emphasize the importance of the theoretical uncertainty related to model assumptions (e.g., envelope structure, and composition of the planetary layers), which may overcome the observational uncertainties soon. We therefore suggest that, along with the great efforts to improve the data, similar efforts should be made on the theoretical front.  

\begin{acknowledgements}
    We thank the referee for valuable comments which significantly improved our paper. C.D. acknowledges support from the Swiss National Science Foundation under grant PZ00P2\_174028. J.H. acknowledges the support from the Swiss National Science Foundation (SNSF) under grant 200020\_172746. This work has been carried out within the frame of the National Center for Competence in Research PlanetS supported by the SNSF.
\end{acknowledgements}

\bibliography{bibliography}

\end{document}